\documentstyle[sprocl]{article}

\begin{document}

\title{PREDICTIONS OF INFLATION: 
THE SLOW ROLL APPROXIMATION VS.~EXACT RESULTS}

\author{Dominik J. SCHWARZ}

\address{Institut f\"ur Theoretische Physik, Robert-Mayer-Stra\ss e 8 -- 10,
Postfach 11 19 32,\\
60054 Frankfurt am Main, Germany\\
E-mail: dschwarz@th.physik.uni-frankfurt.de} 

\author{J\'er\^ome MARTIN}

\address{DARC, Observatoire de Paris, UPR 176 CNRS,\\
92195 Meudon Cedex, France\\
E-mail: martin@edelweiss.obspm.fr}

\maketitle\abstracts{
Inflation predicts the generation of cosmological perturbations.
Usually, the power spectra for the scalar and tensor modes are calculated 
with help of the slow roll approximation. In the case of power law 
inflation an exact result is available. We compare the predictions
for the cosmic microwave background anisotropies from the slow roll 
approximation with the exact results from power law 
inflation. We find that the so-called consistency check from the 
slow roll approximation, $C_2^{\rm T}/C_2^{\rm S} \approx -6.93 n_{\rm T}$,
may differ considerably from the exact result.}

\section{Introduction}

The inflationary scenario allows to solve the horizon and flatness
problems and predicts the generation of density (scalar) 
perturbations and of gravitational waves (tensor perturbations) \cite{MFB}. 
Due to the Sachs-Wolfe effect \cite{SW} those perturbations can be 
observed in the cosmic microwave background (CMB). 
The COBE satellite has measured 
these anisotropies for the first time \cite{COBE}. Forthcoming high 
precision observations, especially the MAP and PLANCK satellites \cite{MP}, 
will determine the temperature correlations with a precision of a few 
percent. Therefore predictions from inflationary models should be
made on the few percent level as well. 

Almost all (analytical) predictions for perturbation spectra from inflation
rely on the slow roll approximation \cite{S,L}. So far, no systematic, 
quantitative analysis on the error of the slow roll approximation has been
performed, neither for the power spectra, nor for the temperature two-point 
correlations. We consider this work as a first step in such an error analysis.
We compare the results from slow roll inflation, i.e.~$a(t) \sim \exp(H t), 
H \sim $ const, with the exact solutions from power law inflation, i.e. 
$a(t) \sim t^p, p=$const. 

In a previous work \cite{MS}, in response to contrary claims,
we showed that the contribution of tensor perturbations with respect to
scalar perturbations to the CMB anisotropies is small for the equation 
of state $\rho \approx - p$ during inflation. However, inflation 
occurs already if $\rho + 3 p < 0$, which does not necessarily lead to 
slow roll inflation.

Power law inflation provides exact solutions for the time evolution of 
cosmological perturbations and inflation can occur 
although the slow roll conditions are violated. It is therefore interesting 
to investigate the difference of the exact predictions with the 
slow roll predictions.

Here, we concentrate on the so-called consistency check, which relates 
the scalar and tensor CMB quadrupole.
 
\section{Observables}

The observable quantities are the temperature two-point correlation
functions, respectively their momenta $C_l^{\rm S,T}$. We define them 
through:
\begin{equation}
\langle \left(\frac{{\rm \delta}T}{T}\right)^{\rm S,T}(\vec{e}_1) 
        \left(\frac{{\rm \delta}T}{T}\right)^{\rm S,T}(\vec{e}_2)\rangle =
\frac 1{4\pi} \sum_l (2l+1) C_l^{\rm S,T} P_l(\cos\delta) \ ,
\end{equation}
where $\vec{e_a}$ denotes the beam direction. The temperature fluctuations 
${\rm \delta }T/T(\vec{e})$ are related to the 
primordial cosmological perturbations by the Sachs-Wolfe 
effect. We evaluated the scalar fluctuations for purely adiabatic 
perturbations (i.e. no entropy perturbations) and for large scale modes, 
such that the sudden decoupling approximation applies. We also assumed that 
the perfect fluid approximation holds.

If the perturbations are of quantum-mechanical origin then the power spectra 
of the Bardeen potential and of the tensor fluctuations are respectively 
given by $P_{\Phi}(k) = 
A_{\rm S} k^{n_{\rm S} - 4}$ and $P_{h}(k) = 
A_{\rm T} k^{n_{\rm T}-3}$. They are related to the $C_l$'s by \cite{PS}:
\begin{eqnarray}
\label{num1}
C_l^{\rm S} &=& \frac{4\pi} 9 \int_0^\infty {{\rm d} k\over k}
[j_l(k\eta_0)]^2 A_{\rm S}k^{n_{\rm S} - 1}, \\
\label{num2}
C_l^{\rm T} &=& \frac{9 \pi}{4} (l-1)l(l+1)(l+2)
\int_0^\infty {{\rm d} k\over k} I_l^2(k\eta_0) A_{\rm T} k^{n_{\rm T}},
\end{eqnarray}
where,
\begin{equation} 
I_l(k\eta_0) \equiv  \int_{k\eta_{\rm e}}^{k\eta_0} 
{j_2(x)j_l(k\eta_0 - x)\over
x (k\eta_0 - x)^2} {\rm d} x,
\end{equation}
$j_l$ being a spherical Bessel function and $\eta _0 $ the time of 
reception of CMB photons.

\section{Slow roll inflation}

Slow roll inflation (at leading order) is 
controlled by three parameters:
\begin{eqnarray}
\epsilon &\equiv & 3 \dot{\varphi}^2/2
(\dot{\varphi}^2/2 + V)^{-1} = -\dot{H}/H^2, \\
\delta &\equiv & -\ddot{\varphi }/(H\dot{\varphi }) =
- \dot{\epsilon }/(2 H \epsilon) + \epsilon ,\\
\xi &\equiv & (\dot{\epsilon }-\dot{\delta })/H \ .
\end{eqnarray}
The universe is inflating as soon 
as $\epsilon <1$. The slow roll 
approximation holds true for $\epsilon \ll 1, 
\delta \ll 1$, and $\xi = {\cal O}(\epsilon^2, \delta^2, \epsilon\delta)$.
There are important examples that do not satisfy all three conditions, e.g.
inflation with a Coleman-Weinberg potential.

At the leading order we find \cite{L,SL}
\begin{equation}
\label{sr}
A_{\rm S}k^{n_{\rm S} - 1} = \frac{9}{25 \pi \epsilon}
\left. \frac{H^2}{m_{\rm Pl}^2} \right|_{k=aH} \ ,\qquad
A_{\rm T}k^{n_{\rm T}} = \frac{16}{\pi}
\left. \frac{H^2}{m_{\rm Pl}^2} \right|_{k=aH} \ .
\end{equation}
With $n_{\rm T} \approx -2\epsilon$ the 
so-called consistency check follows:
\begin{equation}
\label{TSsr}
C_2^{\rm T}/C_2^{\rm S} \approx -6.93 n_{\rm T} \ .
\end{equation}
The integrations in (\ref{num2}) have been performed 
numerically at $n_{\rm T}= 0$ and expressions in Eq. (\ref{num1}) have 
been evaluated for $n_{\rm S}=1$. 

For $\epsilon \to 0$ (no roll), $A_{\rm S}$ diverges, which means that 
linear perturbation theory does not apply for very small values of $\epsilon$. 
Thus, nowhere in the parameter space $(\epsilon, \delta, \xi)$ the slow roll 
approximation is exact. Including higher orders in $\epsilon, \delta,
\xi$ and/or introducing new parameters does not change this conclusion.

\section{Power law inflation}

This special model is equivalent to scalar field inflation with the
potential $V(\varphi) = V_0 \exp(\pm 4\sqrt{\pi}l_{\rm Pl} \varphi/\sqrt{p})$.
For power law inflation $\epsilon = \delta = 1/p$, $\xi = 0$. Thus, 
$p > 1$ is sufficient for inflation. In the limit $p \gg 1$ power law 
inflation and slow roll inflation with $\epsilon = \delta$ agree 
at the leading order in $\epsilon$. 

The exact evolution of the cosmological fluctuations for power law inflation
[$a(\eta) = l_0 |\eta|^{\beta+1}; p=(\beta+1)/(\beta+2)$] gives rise to 
\cite{MS,SL}
\begin{equation}
A_{\rm S} = {l_{\rm Pl}^2\over l_0^2} {9\over 25 \pi \epsilon} f(\beta)\ ,
\quad n_{\rm S} = 2 \beta + 5
\end{equation}
with $f(\beta) = 4\pi/
[2^{\beta+2}\cos(\beta\pi) \Gamma(\beta + 3/2)]^2$ and 
\begin{equation}
A_{\rm T} = {l_{\rm Pl}^2\over l_0^2} {16 \over \pi}  f(\beta)\ ,
\quad n_{\rm T} = 2 \beta + 4 \ .
\end{equation}
Now, the ratio of tensor to scalar contributions to the temperature 
fluctuations is given by: 
\begin{equation}
\label{TSpl}
C_2^{\rm T}/C_2^{\rm S} \approx -6.93 F(n_{\rm T})
n_{\rm T}/(1 - n_{\rm T}/2) \ ,  
\end{equation}
where $F(n_{\rm T})$ denotes a numerical integration, which
is normalized such that $F(0)=1$. We have used the relation
$\epsilon = - n_{\rm T}/(2 - n_{\rm T})$ and have put all other 
dependence on $n_{\rm T}$ and on $n_{\rm S} (= n_{\rm T} + 1)$ into the 
function $F(n_{\rm T})$.

\section{Discussion and conclusion}

The exact result and the slow roll
result differ by a factor $F(n_{\rm T})/(1 - n_{\rm T}/2)$. For inflation 
$p > 1$, which translates into $0 > n_{\rm T} = - 2/(p-1) > - \infty$ we find 
that the slow roll result might differ considerably, e.g.~for $p=2$ the
error is a factor $F(-2)/2 \approx 0.34$. Only 
for $p > 100$, i.e. for $1>n_{\rm S}>0.99$, the error 
of the slow roll approximation is less than $1 \%$. 

For small numbers $l$ the cosmic variance introduces an uncertainty of
$\Delta C_l = \sqrt{2/2l+1} C_l$, for the quadrupole $\Delta C_2 = 0.63 C_2$. 
Thus, even at the largest scales the error from the slow roll approximation 
might be as big as the cosmic variance. For intermediate values $l < 30$  
the error from the slow roll approximation is even more important.

Our main conclusion is that the consistency check, Eq.~(\ref{TSsr}),
is not valid for an arbitrary inflationary model. When the slow roll 
approximation does not apply, as for power law inflation, we expect 
significant modifications.

\section*{Acknowledgments}
D.J.S. is supported by an Alexander von Humboldt fellowship.


\section*{References}
\begin{thebibliography}{99}
\bibitem{MFB}V. F. Mukhanov, H. A. Feldman, and R. H. Brandenberger, Phys.
         Rep. {\bf 215}, 203 (1992).
\bibitem{SW} R. K. Sachs and A. M. Wolfe, Ap. J. {\bf 147}, 73 (1967).
\bibitem{COBE}G. Smoot et al., Ap. J. {\bf 396}, L1 (1992).
\bibitem{MP} See http://astro.estec.esa.nl/SA-general/Projects/Planck/
and\\ http://map.gsfc.nasa.gov/.
\bibitem{S} P. J. Steinhardt, Int. J. Mod. Phys. A {\bf 10}, 1091 (1995).
\bibitem{L} J. E. Lidsey et al., Rev. Mod. Phys. {\bf 69}, 373 (1997).
\bibitem{MS}J. Martin and D.J. Schwarz, Phys. Rev. D {\bf 57}, 3302 (1998).
\bibitem{PS}P. J. E. Peebles, Ap. J. {\bf 263}, L1 (1982); A. A. Starobinskii,
Pis'ma Astron. Zh. {\bf 11}, 323 (1985) 
[Sov. Astron. Lett. {\bf 11}, 133 (1985)].
\bibitem{SL} E. D. Stewart and D. H. Lyth, Phys. Lett. B {\bf 302}, 171 (1993).
\end{thebibliography}
\end{document}